\documentclass{article}

\PassOptionsToPackage{numbers, compress}{natbib}

    \usepackage[final]{neurips_2022}

\usepackage[utf8]{inputenc} 
\usepackage[T1]{fontenc}    
\usepackage{hyperref}       
\usepackage{url}            
\usepackage{booktabs}       
\usepackage{amsfonts}       
\usepackage{nicefrac}       
\usepackage{microtype}      
\usepackage{xcolor}         

\usepackage{graphicx}
\usepackage{wrapfig}

\newcommand{\beginsupplement}{%
    \setcounter{table}{0}
    \renewcommand{\thetable}{S\arabic{table}}%
    \setcounter{figure}{0}
    \renewcommand{\thefigure}{S\arabic{figure}}%
}

\title{A single-cell gene expression language model}

\author{%
  William Connell \\
  Department of Pharmaceutical Chemistry\\
  Institute for Neurodegenerative Diseases\\
  University of California, San Francisco\\
  San Francisco, CA 94143 \\
  \texttt{connell@keiserlab.org} \\
\And
  Umair Khan \\
  Department of Pharmaceutical Chemistry\\
  Institute for Neurodegenerative Diseases\\
  University of California, San Francisco\\
  San Francisco, CA 94143 \\
  \texttt{ukhan@keiserlab.org} \\
\And
  Michael J. Keiser \\
  Department of Pharmaceutical Chemistry\\
  Institute for Neurodegenerative Diseases\\
  University of California, San Francisco\\
  San Francisco, CA 94143 \\
  \texttt{keiser@keiserlab.org} \\
 }

\begin{document}

\maketitle

\begin{abstract}
Gene regulation is a dynamic process that connects genotype and phenotype. Given the difficulty of physically mapping mammalian gene circuitry, we require new computational methods to learn regulatory rules. Natural language is a valuable analogy to the communication of regulatory control. Machine learning systems model natural language by explicitly learning context dependencies between words. We propose a similar system applied to single-cell RNA expression profiles to learn context dependencies between genes. Our model, Exceiver, is trained across a diversity of cell types using a self-supervised task formulated for discrete count data, accounting for feature sparsity. We found agreement between the similarity profiles of latent sample representations and learned gene embeddings with respect to biological annotations. We evaluated Exceiver on a new dataset and a downstream prediction task and found that pretraining supports transfer learning. Our work provides a framework to model gene regulation on a single-cell level and transfer knowledge to downstream tasks.
\end{abstract}

\section{Introduction}
\label{intro}

Many biological processes regulate the relationship between genotype and phenotype. On one hand, classical genetics defines simple hereditary rules. On the other hand, complex regulatory networks mediate response to the environment. Eventually, we may hope to model molecular circuitry comprehensively to accurately predict phenotypes. In this direction, learned generalizations of biological processes may support medical interventions such as individual disease risk prediction and patient therapy selection.

Large-scale cellular assays capture snapshots of complex and dynamic biological processes such as gene regulation, with RNA abundances being one measurable outcome. Single-cell RNA sequencing (scRNA-seq) observations can relate cellular states and mRNA expression relationships, revealing gene programs corresponding to disease processes, genetic perturbations, and therapeutic interventions \citep{Hwang2018-wz}. Given the difficulty of physically mapping regulatory circuitry explicitly, we hypothesized a model trained on a large volume of transcriptomic profiles would instead implicitly learn RNA expression dependencies that reflect regulatory logic. 

Pretrained models in natural language processing, computer vision, and protein modeling motivate a similar approach in systems biology \citep{Brown2020-mu, He2016-qs, Rives2021-ou}. Pretrained models that transfer to downstream tasks share three components leveraging domain-specific inductive biases. First, sufficient unlabeled data volumes provide enough information for highly parameterized models to learn complex relationships between features. Second, models learn from unlabeled data in a self-supervised manner, often by feature masking, in which unmasked features are used to predict a fraction of values that are masked. Third, an attention mechanism learns the dependencies between features. Traditionally, a transformer applies self-attention to learn context-dependent feature representations. Given the success of this recipe across various domains, we propose to model gene regulation similarly.
	
Building on sequence modeling, Exceiver (Expression-Perceiver) is a single-cell gene expression language model pretrained on an atlas of transcriptomic data. We leveraged the Perceiver IO framework to train a long-context sequence model on all protein-coding genes in a self-supervised manner \citep{Jaegle2021-fv, Jaegle2021-tl}. We evaluated latent sample representations with respect to metadata labels including cell compartment, tissue, and cell ontology. We analyzed the similarity of learned gene embeddings relative to known molecular interactions. Finally, we assessed pretrained Exceiver models on new datasets and a downstream task. Exceiver provides a framework to learn gene regulatory logic from unlabeled single-cell transcriptomes and transfer knowledge to new domains.

\begin{figure}
  \centering
  \includegraphics[width=\textwidth]{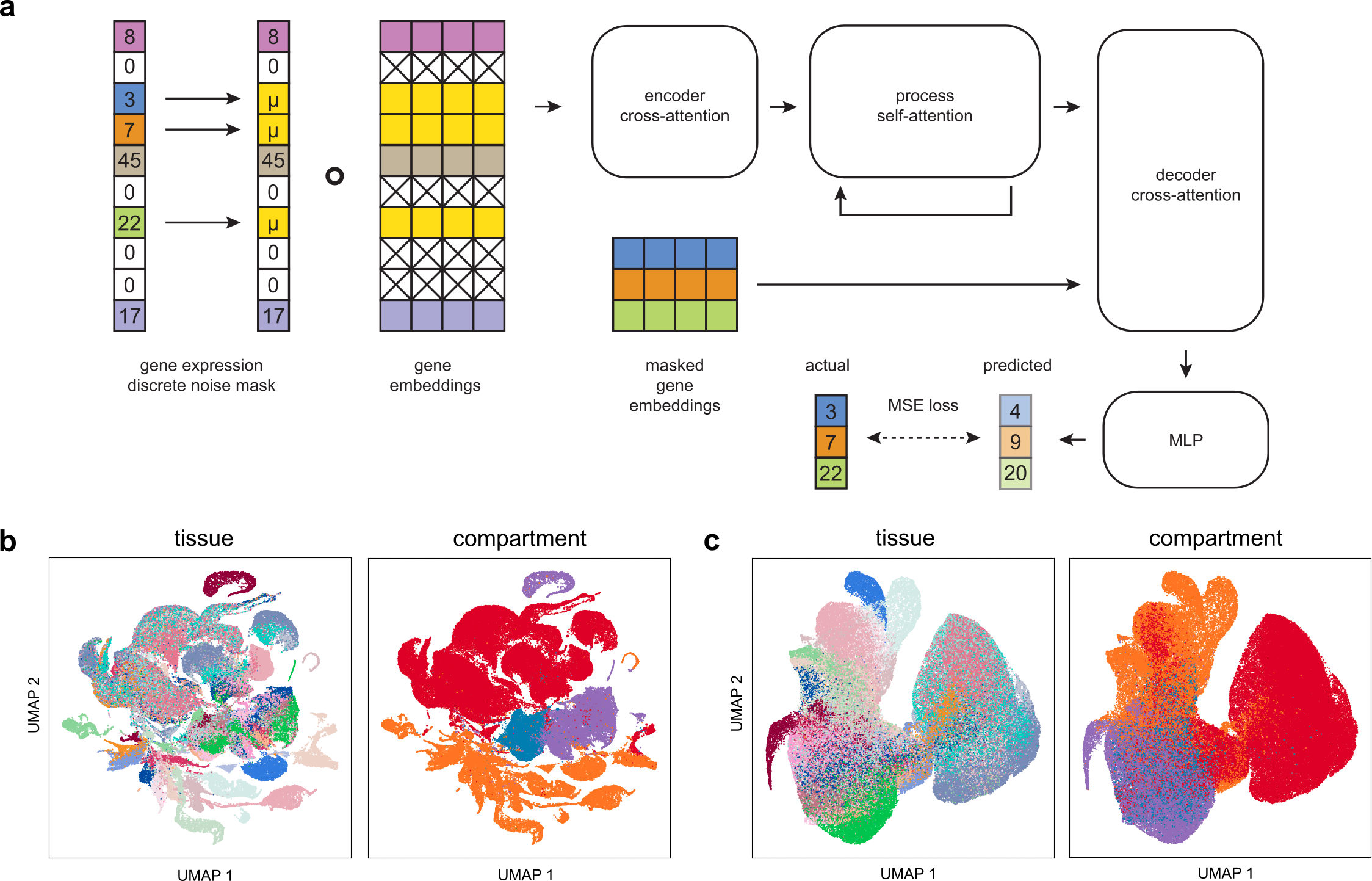}
  \caption{\textbf{Exceiver learns cell embeddings reflecting tissue and compartment.} (a) Architectural overview and pretraining strategy. UMAP of (b) original data PCA embeddings and (c) Exceiver sample embeddings colored by tissue type and compartment.}
  \label{fig-1}
\end{figure}

\section{Exceiver accounts for discrete features and technical dropout in scRNA-seq self-supervised pretraining}
\label{arch}

Exceiver builds on Perceiver IO to encode single-cell transcriptomic profiles. Perceiver IO scales linearly with the size of inputs and outputs, allowing tractable transformer-based encoding and decoding of long-context sequences. Exceiver retains the core Perceiver IO architectural components: a cross-attention encoder, a self-attention latent process module, and a cross-attention decoder (Figure 1a; Methods). 

Exceiver extends this general architecture to accommodate various biological and experimental priors. To account for discrete RNA abundances, Exceiver represents individual genes as learnable embeddings. Global gene embedding vectors are scaled by expression values to incorporate observed RNA counts. Exceiver can also augment the learning processes with prior knowledge through the integration of experimental metadata. An auxiliary classification option encodes sample class labels, such as tissue type, as an additional embedding. In addition to biological priors, the Exceiver framework acknowledges that technical dropout influences scRNA-seq experimental measurements. Exceiver masks attention computation at these values to mitigate feature sparsity biases in the learning process (Figure 1a; Methods). 

To leverage unlabeled single-cell atlases for pretraining, we propose discrete noise masking (DNM), a novel self-supervised task for count-based RNA abundances. DNM randomly chooses a fraction of genes to mask each time a cell is sampled for training. A mask embedding replaces sampled gene embeddings and a noised expression value replaces the true expression value. In our experiments DNM simply samples the mean of the feature distribution for expression noising; however, DNM may extend to other distributions. (Figure 1a; Methods).

\section{Exceiver learns representations that reflect known biology}
\label{known_bio}

We trained Exceiver on the Tabula Sapiens, a healthy human atlas of approximately 500,000 single cells from 24 organs of 15 individuals \citep{Tabula_Sapiens_Consortium2022-tm}. We randomly split the data into 70\% training and 30\% validation sets and pretrained Exceiver with DNM and tissue identity as an auxiliary task. The model converged to an explained variance (EV) of approximately 0.73 and a classification accuracy of 0.61 on the validation dataset. 

To assess whether Exceiver learned sample representations that reflect biological relationships, we evaluated latent representation similarity relative to that of metadata annotations. Qualitatively, both the original data and Exceiver latent representations reflected tissue of origin and cell compartment (evolutionary lineage) (Figure 1b,c). We applied k-means clustering to samples and computed the adjusted mutual information (AMI) between derived cluster labels and true labels. Exceiver latent representations achieved a considerably higher AMI than original samples by tissue, donor, and compartment labels (Table 1). Given its role as the auxiliary classification task, we expected latent representations to cluster by tissue. However, structure also increased for donor and compartment labels. Cell ontology, a fine-grained label, saw decreased clustering relative to original samples. Exceiver’s learned sample representations reflected known relationships and led us to interrogate similar structure in learned gene embeddings.

\begin{table}
  \centering
  \begin{tabular}{lll}
    \toprule
    sample label & original samples & latent representations \\
    \midrule
    tissue & 0.25 & \textbf{0.42}\\
    method & 0.00 & 0.00 \\
    donor & 0.08 & \textbf{0.16} \\
    cell ontology & \textbf{0.56} & 0.43 \\
    compartment & 0.07 & \textbf{0.44}\\
    gender & 0.01 & 0.00 \\
    \bottomrule\\
  \end{tabular}
  \caption{\textbf{Latent sample representations reflect biological annotations.} AMI was computed between k-means clustering derived labels and true labels of original sample embeddings and Exceiver latent sample representations.}
  \label{table-1}
\end{table}

We hypothesized that similarity profiles of learned gene embeddings may reflect known gene associations. To analyze learned gene relationships, we extracted the vocabulary of global gene embeddings and applied unsupervised Leiden clustering (Figure 2a). Then, we queried the STRING database with the gene list from each cluster and calculated network enrichment \citep{Szklarczyk2015-dk}. 66\% (39/59) of Exceiver gene clusters had more interactions than expected by chance (Figure 2b). Additionally, we investigated functional gene associations through Gene Ontology (GO). For example, cluster 17 reflected significant enrichment of GO process terms associated with muscle function (Figure 2c, Figure S1a). In another case, an extremely enriched cluster contained a large portion of ribosomal genes (Figure S1b). Overall, Exceiver gene embeddings reflected network biology associations and captured functional gene relationships.

\begin{figure}
  \centering
  \includegraphics[width=\textwidth]{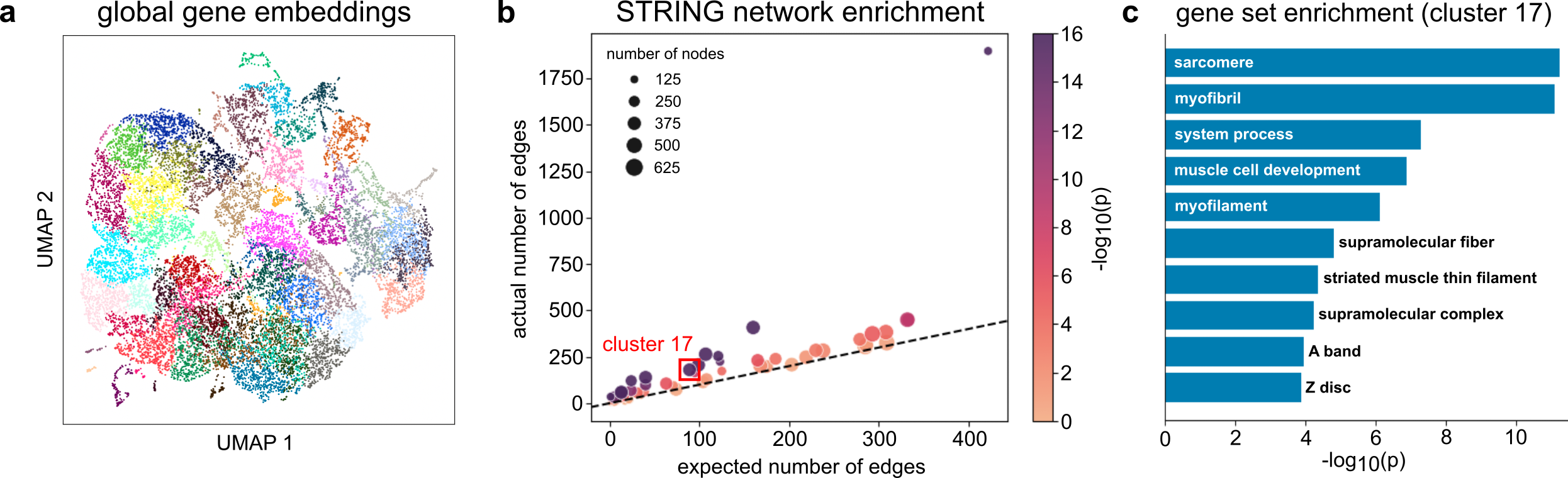}
  \caption{\textbf{Learned gene embedding similarity profiles reflect network biology.} (a) UMAP of global gene embeddings colored by Leiden cluster. (b) STRING network enrichment plot of gene clusters. (c) Gene set enrichment analysis of cluster 17.}
  \label{fig-2}
\end{figure}

\section{Pretrained Exceiver transfers to new datasets and predicts drug response}
\label{transfer}

\begin{wrapfigure}{l}{0.4\textwidth}
    \includegraphics[width=0.4\textwidth]{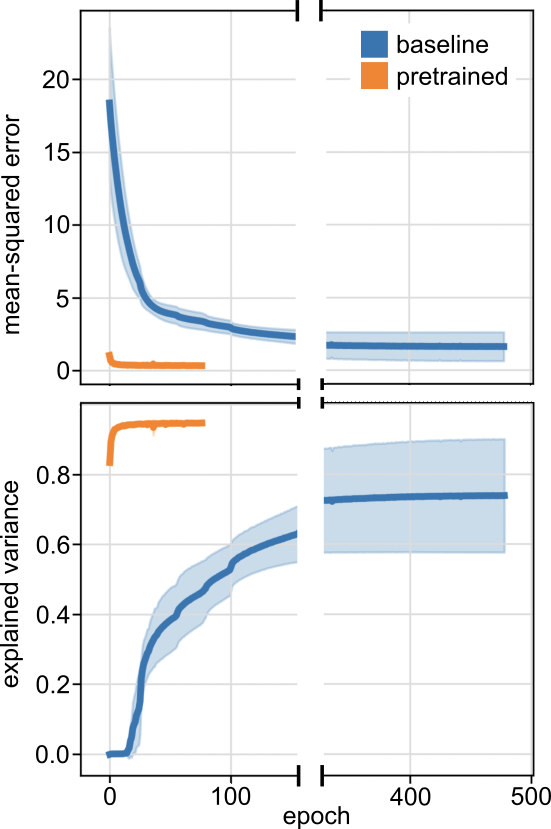}
    \caption{\textbf{Pretrained Exceiver encodes a new dataset.} Validation loss and explained variance curves of baseline and pretrained Exceiver models. Models were trained and evaluated with DNM on Bi et al. data.}
    \vspace{-9pt}
    \label{fig-3}
\end{wrapfigure}

Next, we evaluated the transferability of a pretrained Exceiver model to a new dataset. We expected that, compared to an untrained Exceiver model, a pretrained model would more rapidly converge to higher performance on a new dataset. We tested this on a scRNA-seq dataset generated at a different time, in a different lab, of a novel disease physiology. Bi et al. investigated tumor and immune cell reprogramming of patients treated with immune checkpoint blockade for metastatic renal cell carcinoma \citep{Bi2021-we}. Biopsies from 8 patients were characterized by scRNA-seq for 35,000 cells. We trained five randomly-initialized Exceiver models on this dataset with DNM and no auxiliary task. We likewise fine-tuned five pretrained Exceiver models. Prior to fine-tuning, the pretrained Exceiver models predicted masked expression values with an average EV of 0.52. The models then converged to an average EV of 0.94 in under 10 epochs. This is in contrast to the baseline models, which converged to an average EV of 0.73 over a 30x longer training interval (~350 epochs) (Figure 3). An Exceiver model pretrained on the self-supervised DNM task learns information that is sufficiently general to apply across new datasets and biological contexts.


Additionally, we hypothesized that a pretrained Exceiver model would support transfer learning to a new downstream task. We turned to a MIX-Seq dataset of pooled cell lines that were transcriptionally characterized after drug treatments. As in the original study, we matched drug response (quantified by area under the dose-response curve) from the Genomics of Drug Sensitivity in Cancer Project (GDSC) screen to MIX-Seq transcriptional profiles \citep{McFarland2020-pd, Yang2013-hc}. We trained five baseline and pretrained Exceiver models on each drug-treated cell line pool (Methods). Consistent with the original study, neither approach learned a relationship between post-perturbation transcriptional profiles and drug responses for navitoclax, taselisib, bortezomib, or gemcitabine, likely due to matching drug responses from an entirely different dataset. By contrast, trametinib and dabrafenib treated cell line pools matched the most samples and here the pretrained Exceiver succeeded whereas the baseline Exceiver did not ($p < 0.05$, Mann-Whitney U; Table 2). Given the high training variances of the baseline models, further optimization may be warranted, but it is clear that the pretrained Exceiver model was the more robust learner in this challenging scenario.

\begin{table}
  \centering
  \begin{tabular}{lllll}
    \toprule
    drug & $n$ samples & baseline (EV) & pretrained (EV) & $p$-value \\
    \midrule
    dabrafenib & 6744 & -0.46 $\pm$ 0.09 & 0.49 $\pm$ 0.18 & 0.008 \\
    trametinib & 6696 & -0.21 $\pm$ 0.15 & 0.14 $\pm$ 0.01 & 0.008 \\
    navitoclax & 5910 & -1.03 $\pm$ 0.40 & -0.57 $\pm$ 0.15 & 0.310 \\
    taselisib & 1327 & 0.004 $\pm$ 0.004 & 0.03 $\pm$ 0.05 & 0.841 \\
    bortezomib & 913 & 0.02 $\pm$ 0.01 & 0.06 $\pm$ 0.07 & 0.222 \\
    gemcitabine & 736 & 0.02 $\pm$ 0.01 & 0.02 $\pm$ 0.05 & 0.421 \\
    \bottomrule\\
  \end{tabular}
  \caption{\textbf{Pretrained Exceiver predicts drug response.} Explained variance (EV) and standard error of baseline and pretrained drug response prediction Exceiver models. MIX-Seq post-perturbation transcriptomic profiles of pooled single-cell drug screens were trained to predict AUC from GDSC screens. (AUC, area under the curve; GDSC, Genomics of Drug Sensitivity in Cancer Project).}
  \label{table-2}
\end{table}

\section{Discussion}

We present Exceiver, a single-cell gene expression language model, whose attention-based transformer backbone encodes long-context transcriptomic profiles. We introduce discrete noise masking, a procedure that masks expression values and enables self-supervised learning on unlabeled, continuously-valued datasets. We show that an Exceiver model trained on the Tabula Sapiens with a self-supervised task learns low dimensional representations that reflect sample annotations. Moreover, learned gene embedding similarity reflects molecular network interactions and functional associations. Finally, we find that a pretrained Exceiver model transfers to new datasets and a drug response prediction task. Exceiver provides a framework to leverage publicly available scRNA-seq datasets and learn robust gene regulatory logic across diverse biological contexts. Exceiver may provide utility in transferring systems knowledge to downstream tasks, from the interrogation of molecular functions to the prediction of comprehensive phenotypes.

\bibliography{exceiver-v1}

\begin{thebibliography}{11}
\providecommand{\natexlab}[1]{#1}
\providecommand{\url}[1]{\texttt{#1}}
\expandafter\ifx\csname urlstyle\endcsname\relax
  \providecommand{\doi}[1]{doi: #1}\else
  \providecommand{\doi}{doi: \begingroup \urlstyle{rm}\Url}\fi

\bibitem[Bi et~al.(2021)Bi, He, Bakouny, Kanodia, Napolitano, Wu, Grimaldi,
  Braun, Cuoco, Mayorga, DelloStritto, Bouchard, Steinharter, Tewari, Vokes,
  Shannon, Sun, Park, Chang, McGregor, Haq, Denize, Signoretti, Guerriero,
  Vigneau, Rozenblatt-Rosen, Rotem, Regev, Choueiri, and Van~Allen]{Bi2021-we}
K.~Bi, M.~X. He, Z.~Bakouny, A.~Kanodia, S.~Napolitano, J.~Wu, G.~Grimaldi,
  D.~A. Braun, M.~S. Cuoco, A.~Mayorga, L.~DelloStritto, G.~Bouchard,
  J.~Steinharter, A.~K. Tewari, N.~I. Vokes, E.~Shannon, M.~Sun, J.~Park, S.~L.
  Chang, B.~A. McGregor, R.~Haq, T.~Denize, S.~Signoretti, J.~L. Guerriero,
  S.~Vigneau, O.~Rozenblatt-Rosen, A.~Rotem, A.~Regev, T.~K. Choueiri, and
  E.~M. Van~Allen.
\newblock Tumor and immune reprogramming during immunotherapy in advanced renal
  cell carcinoma.
\newblock \emph{Cancer Cell}, 39\penalty0 (5):\penalty0 649--661.e5, May 2021.
\newblock ISSN 1535-6108, 1878-3686.
\newblock \doi{10.1016/j.ccell.2021.02.015}.
\newblock URL \url{http://dx.doi.org/10.1016/j.ccell.2021.02.015}.

\bibitem[Brown et~al.(2020)Brown, Mann, Ryder, Subbiah, Kaplan, Dhariwal,
  Neelakantan, Shyam, Sastry, Askell, Agarwal, Herbert-Voss, Krueger, Henighan,
  Child, Ramesh, Ziegler, Wu, Winter, Hesse, Chen, Sigler, Litwin, Gray, Chess,
  Clark, Berner, McCandlish, Radford, Sutskever, and Amodei]{Brown2020-mu}
T.~B. Brown, B.~Mann, N.~Ryder, M.~Subbiah, J.~Kaplan, P.~Dhariwal,
  A.~Neelakantan, P.~Shyam, G.~Sastry, A.~Askell, S.~Agarwal, A.~Herbert-Voss,
  G.~Krueger, T.~Henighan, R.~Child, A.~Ramesh, D.~M. Ziegler, J.~Wu,
  C.~Winter, C.~Hesse, M.~Chen, E.~Sigler, M.~Litwin, S.~Gray, B.~Chess,
  J.~Clark, C.~Berner, S.~McCandlish, A.~Radford, I.~Sutskever, and D.~Amodei.
\newblock Language models are {Few-Shot} learners.
\newblock In \emph{Adv. Neural Inf. Process. Syst.}, volume~33, pages
  1877--1901, May 2020.
\newblock URL
  \url{https://papers.nips.cc/paper/2020/file/1457c0d6bfcb4967418bfb8ac142f64a-Paper.pdf}.

\bibitem[He et~al.(2016)He, Zhang, Ren, and Sun]{He2016-qs}
K.~He, X.~Zhang, S.~Ren, and J.~Sun.
\newblock Deep residual learning for image recognition.
\newblock In \emph{Proceedings of the {IEEE} Conference on Computer Vision and
  Pattern Recognition}, June 2016.
\newblock URL
  \url{https://openaccess.thecvf.com/content_cvpr_2016/papers/He_Deep_Residual_Learning_CVPR_2016_paper.pdf}.

\bibitem[Hwang et~al.(2018)Hwang, Lee, and Bang]{Hwang2018-wz}
B.~Hwang, J.~H. Lee, and D.~Bang.
\newblock Single-cell {RNA} sequencing technologies and bioinformatics
  pipelines.
\newblock \emph{Exp. Mol. Med.}, 50\penalty0 (8):\penalty0 1--14, Aug. 2018.
\newblock ISSN 1226-3613, 2092-6413.
\newblock \doi{10.1038/s12276-018-0071-8}.
\newblock URL \url{http://dx.doi.org/10.1038/s12276-018-0071-8}.

\bibitem[Jaegle et~al.(2021{\natexlab{a}})Jaegle, Borgeaud, Alayrac, Doersch,
  Ionescu, Ding, Koppula, Zoran, Brock, Shelhamer, H{\'e}naff, Botvinick,
  Zisserman, Vinyals, and Carreira]{Jaegle2021-fv}
A.~Jaegle, S.~Borgeaud, J.-B. Alayrac, C.~Doersch, C.~Ionescu, D.~Ding,
  S.~Koppula, D.~Zoran, A.~Brock, E.~Shelhamer, O.~H{\'e}naff, M.~M. Botvinick,
  A.~Zisserman, O.~Vinyals, and J.~Carreira.
\newblock Perceiver {IO}: A general architecture for structured inputs \&
  outputs.
\newblock July 2021{\natexlab{a}}.
\newblock URL \url{http://arxiv.org/abs/2107.14795}.

\bibitem[Jaegle et~al.(2021{\natexlab{b}})Jaegle, Gimeno, Brock, Zisserman,
  Vinyals, and Carreira]{Jaegle2021-tl}
A.~Jaegle, F.~Gimeno, A.~Brock, A.~Zisserman, O.~Vinyals, and J.~Carreira.
\newblock Perceiver: General perception with iterative attention.
\newblock In \emph{Proceedings of Machine Learning Research}, volume~38, Mar.
  2021{\natexlab{b}}.
\newblock URL \url{http://proceedings.mlr.press/v139/jaegle21a/jaegle21a.pdf}.

\bibitem[McFarland et~al.(2020)McFarland, Paolella, Warren, Geiger-Schuller,
  Shibue, Rothberg, Kuksenko, Colgan, Jones, Chambers, Dionne, Bender, Wolpin,
  Ghandi, Tirosh, Rozenblatt-Rosen, Roth, Golub, Regev, Aguirre, Vazquez, and
  Tsherniak]{McFarland2020-pd}
J.~M. McFarland, B.~R. Paolella, A.~Warren, K.~Geiger-Schuller, T.~Shibue,
  M.~Rothberg, O.~Kuksenko, W.~N. Colgan, A.~Jones, E.~Chambers, D.~Dionne,
  S.~Bender, B.~M. Wolpin, M.~Ghandi, I.~Tirosh, O.~Rozenblatt-Rosen, J.~A.
  Roth, T.~R. Golub, A.~Regev, A.~J. Aguirre, F.~Vazquez, and A.~Tsherniak.
\newblock Multiplexed single-cell transcriptional response profiling to define
  cancer vulnerabilities and therapeutic mechanism of action.
\newblock \emph{Nat. Commun.}, 11\penalty0 (1):\penalty0 1--15, Aug. 2020.
\newblock ISSN 2041-1723, 2041-1723.
\newblock \doi{10.1038/s41467-020-17440-w}.
\newblock URL \url{https://www.nature.com/articles/s41467-020-17440-w}.

\bibitem[Rives et~al.(2021)Rives, Meier, Sercu, Goyal, Lin, Liu, Guo, Ott,
  Lawrence~Zitnick, Ma, and Fergus]{Rives2021-ou}
A.~Rives, J.~Meier, T.~Sercu, S.~Goyal, Z.~Lin, J.~Liu, D.~Guo, M.~Ott,
  C.~Lawrence~Zitnick, J.~Ma, and R.~Fergus.
\newblock Biological structure and function emerge from scaling unsupervised
  learning to 250 million protein sequences.
\newblock \emph{PNAS}, Apr. 2021.
\newblock \doi{10.1101/622803}.
\newblock URL \url{https://github.com/facebookresearch/esm}.

\bibitem[Szklarczyk et~al.(2015)Szklarczyk, Franceschini, Wyder, Forslund,
  Heller, Huerta-Cepas, Simonovic, Roth, Santos, Tsafou, Kuhn, Bork, Jensen,
  and von Mering]{Szklarczyk2015-dk}
D.~Szklarczyk, A.~Franceschini, S.~Wyder, K.~Forslund, D.~Heller,
  J.~Huerta-Cepas, M.~Simonovic, A.~Roth, A.~Santos, K.~P. Tsafou, M.~Kuhn,
  P.~Bork, L.~J. Jensen, and C.~von Mering.
\newblock {STRING} v10: protein-protein interaction networks, integrated over
  the tree of life.
\newblock \emph{Nucleic Acids Res.}, 43\penalty0 (Database issue):\penalty0
  D447--52, Jan. 2015.
\newblock ISSN 0305-1048, 1362-4962.
\newblock \doi{10.1093/nar/gku1003}.
\newblock URL \url{http://dx.doi.org/10.1093/nar/gku1003}.

\bibitem[{Tabula Sapiens Consortium*} et~al.(2022){Tabula Sapiens Consortium*},
  Jones, Karkanias, Krasnow, Pisco, Quake, Salzman, Yosef, Bulthaup, Brown,
  Harper, Hemenez, Ponnusamy, Salehi, Sanagavarapu, Spallino, Aaron,
  Concepcion, Gardner, Kelly, Neidlinger, Wang, Crasta, Kolluru, Morri, Pisco,
  Tan, Travaglini, Xu, Alc{\'a}ntara-Hern{\'a}ndez, Almanzar, Antony,
  Beyersdorf, Burhan, Calcuttawala, Carter, Chan, Chang, Chang, Colville,
  Crasta, Culver, Cvijovi{\'c}, D'Amato, Ezran, Galdos, Gillich, Goodyer, Hang,
  Hayashi, Houshdaran, Huang, Irwin, Jang, Juanico, Kershner, Kim, Kiss,
  Kolluru, Kong, Kumar, Kuo, Leylek, Li, Loeb, Lu, Mantri, Markovic, McAlpine,
  de~Morree, Morri, Mrouj, Mukherjee, Muser, Neuh{\"o}fer, Nguyen, Perez,
  Phansalkar, Pisco, Puluca, Qi, Rao, Raquer-McKay, Schaum, Scott,
  Seddighzadeh, Segal, Sen, Sikandar, Spencer, Steffes, Subramaniam, Swarup,
  Swift, Travaglini, Van~Treuren, Trimm, Veizades, Vijayakumar, Vo, Vorperian,
  Wang, Weinstein, Winkler, Wu, Xie, Yung, Zhang, Detweiler, Mekonen, Neff,
  Sit, Tan, Yan, Bean, Charu, Forg{\'o}, Martin, Ozawa, Silva, Tan, Toland,
  Vemuri, Afik, Awayan, Botvinnik, Byrne, Chen, Dehghannasiri, Detweiler,
  Gayoso, Granados, Li, Mahmoudabadi, McGeever, de~Morree, Olivieri, Park,
  Pisco, Ravikumar, Salzman, Stanley, Swift, Tan, Tan, Tarashansky, Vanheusden,
  Vorperian, Wang, Wang, Xing, Xu, Yosef, Alc{\'a}ntara-Hern{\'a}ndez, Antony,
  Chan, Chang, Colville, Crasta, Culver, Dethlefsen, Ezran, Gillich, Hang, Ho,
  Irwin, Jang, Kershner, Kong, Kumar, Kuo, Leylek, Liu, Loeb, Lu, Maltzman,
  Metzger, de~Morree, Neuh{\"o}fer, Perez, Phansalkar, Qi, Rao, Raquer-McKay,
  Sasagawa, Scott, Sinha, Song, Spencer, Swarup, Swift, Travaglini, Trimm,
  Veizades, Vijayakumar, Wang, Wang, Winkler, Xie, Yung, Artandi, Beachy,
  Clarke, Giudice, Huang, Huang, Idoyaga, Kim, Krasnow, Kuo, Nguyen, Quake,
  Rando, Red-Horse, Reiter, Relman, Sonnenburg, Wang, Wu, Wu, and
  Wyss-Coray]{Tabula_Sapiens_Consortium2022-tm}
{Tabula Sapiens Consortium*}, R.~C. Jones, J.~Karkanias, M.~A. Krasnow, A.~O.
  Pisco, S.~R. Quake, J.~Salzman, N.~Yosef, B.~Bulthaup, P.~Brown, W.~Harper,
  M.~Hemenez, R.~Ponnusamy, A.~Salehi, B.~A. Sanagavarapu, E.~Spallino, K.~A.
  Aaron, W.~Concepcion, J.~M. Gardner, B.~Kelly, N.~Neidlinger, Z.~Wang,
  S.~Crasta, S.~Kolluru, M.~Morri, A.~O. Pisco, S.~Y. Tan, K.~J. Travaglini,
  C.~Xu, M.~Alc{\'a}ntara-Hern{\'a}ndez, N.~Almanzar, J.~Antony, B.~Beyersdorf,
  D.~Burhan, K.~Calcuttawala, M.~M. Carter, C.~K.~F. Chan, C.~A. Chang,
  S.~Chang, A.~Colville, S.~Crasta, R.~N. Culver, I.~Cvijovi{\'c}, G.~D'Amato,
  C.~Ezran, F.~X. Galdos, A.~Gillich, W.~R. Goodyer, Y.~Hang, A.~Hayashi,
  S.~Houshdaran, X.~Huang, J.~C. Irwin, S.~Jang, J.~V. Juanico, A.~M. Kershner,
  S.~Kim, B.~Kiss, S.~Kolluru, W.~Kong, M.~E. Kumar, A.~H. Kuo, R.~Leylek,
  B.~Li, G.~B. Loeb, W.-J. Lu, S.~Mantri, M.~Markovic, P.~L. McAlpine,
  A.~de~Morree, M.~Morri, K.~Mrouj, S.~Mukherjee, T.~Muser, P.~Neuh{\"o}fer,
  T.~D. Nguyen, K.~Perez, R.~Phansalkar, A.~O. Pisco, N.~Puluca, Z.~Qi, P.~Rao,
  H.~Raquer-McKay, N.~Schaum, B.~Scott, B.~Seddighzadeh, J.~Segal, S.~Sen,
  S.~Sikandar, S.~P. Spencer, L.~C. Steffes, V.~R. Subramaniam, A.~Swarup,
  M.~Swift, K.~J. Travaglini, W.~Van~Treuren, E.~Trimm, S.~Veizades,
  S.~Vijayakumar, K.~C. Vo, S.~K. Vorperian, W.~Wang, H.~N.~W. Weinstein,
  J.~Winkler, T.~T.~H. Wu, J.~Xie, A.~R. Yung, Y.~Zhang, A.~M. Detweiler,
  H.~Mekonen, N.~F. Neff, R.~V. Sit, M.~Tan, J.~Yan, G.~R. Bean, V.~Charu,
  E.~Forg{\'o}, B.~A. Martin, M.~G. Ozawa, O.~Silva, S.~Y. Tan, A.~Toland,
  V.~N.~P. Vemuri, S.~Afik, K.~Awayan, O.~B. Botvinnik, A.~Byrne, M.~Chen,
  R.~Dehghannasiri, A.~M. Detweiler, A.~Gayoso, A.~A. Granados, Q.~Li,
  G.~Mahmoudabadi, A.~McGeever, A.~de~Morree, J.~E. Olivieri, M.~Park, A.~O.
  Pisco, N.~Ravikumar, J.~Salzman, G.~Stanley, M.~Swift, M.~Tan, W.~Tan, A.~J.
  Tarashansky, R.~Vanheusden, S.~K. Vorperian, P.~Wang, S.~Wang, G.~Xing,
  C.~Xu, N.~Yosef, M.~Alc{\'a}ntara-Hern{\'a}ndez, J.~Antony, C.~K.~F. Chan,
  C.~A. Chang, A.~Colville, S.~Crasta, R.~Culver, L.~Dethlefsen, C.~Ezran,
  A.~Gillich, Y.~Hang, P.-Y. Ho, J.~C. Irwin, S.~Jang, A.~M. Kershner, W.~Kong,
  M.~E. Kumar, A.~H. Kuo, R.~Leylek, S.~Liu, G.~B. Loeb, W.-J. Lu, J.~S.
  Maltzman, R.~J. Metzger, A.~de~Morree, P.~Neuh{\"o}fer, K.~Perez,
  R.~Phansalkar, Z.~Qi, P.~Rao, H.~Raquer-McKay, K.~Sasagawa, B.~Scott,
  R.~Sinha, H.~Song, S.~P. Spencer, A.~Swarup, M.~Swift, K.~J. Travaglini,
  E.~Trimm, S.~Veizades, S.~Vijayakumar, B.~Wang, W.~Wang, J.~Winkler, J.~Xie,
  A.~R. Yung, S.~E. Artandi, P.~A. Beachy, M.~F. Clarke, L.~C. Giudice, F.~W.
  Huang, K.~C. Huang, J.~Idoyaga, S.~K. Kim, M.~Krasnow, C.~S. Kuo, P.~Nguyen,
  S.~R. Quake, T.~A. Rando, K.~Red-Horse, J.~Reiter, D.~A. Relman, J.~L.
  Sonnenburg, B.~Wang, A.~Wu, S.~M. Wu, and T.~Wyss-Coray.
\newblock The tabula sapiens: A multiple-organ, single-cell transcriptomic
  atlas of humans.
\newblock \emph{Science}, 376\penalty0 (6594):\penalty0 eabl4896, May 2022.
\newblock ISSN 0036-8075, 1095-9203.
\newblock \doi{10.1126/science.abl4896}.
\newblock URL \url{http://dx.doi.org/10.1126/science.abl4896}.

\bibitem[Yang et~al.(2013)Yang, Soares, Greninger, Edelman, Lightfoot, Forbes,
  Bindal, Beare, Smith, Thompson, Ramaswamy, Futreal, Haber, Stratton, Benes,
  McDermott, and Garnett]{Yang2013-hc}
W.~Yang, J.~Soares, P.~Greninger, E.~J. Edelman, H.~Lightfoot, S.~Forbes,
  N.~Bindal, D.~Beare, J.~A. Smith, I.~R. Thompson, S.~Ramaswamy, P.~A.
  Futreal, D.~A. Haber, M.~R. Stratton, C.~Benes, U.~McDermott, and M.~J.
  Garnett.
\newblock Genomics of drug sensitivity in cancer ({GDSC)}: a resource for
  therapeutic biomarker discovery in cancer cells.
\newblock \emph{Nucleic Acids Res.}, 41\penalty0 (Database issue):\penalty0
  D955--61, Jan. 2013.
\newblock ISSN 0305-1048, 1362-4962.
\newblock \doi{10.1093/nar/gks1111}.
\newblock URL \url{http://dx.doi.org/10.1093/nar/gks1111}.

\end{thebibliography}

\section*{Supplementary Materials}
\label{sup_mat}

\section{Methods}
\label{methods}

\subsection{Model architecture}

Perceiver IO is a general purpose model architecture that adapts to any task with structured input and output. Since self-attention complexity scales quadratically with input size, it cannot be directly applied to high-dimensional data, such as scRNA-seq readouts. Perceiver IO addresses this issue by introducing cross-attentional encoder and decoder mechanisms that project to and from a lower-dimensional latent space where full self-attention can be applied. Each of the blocks in the Perceiver IO architecture are transformer-style modules characterized by query-key-value attention followed by a multilayer perceptron (MLP) and residual connections. For more details please reference the original Perceiver IO manuscript \citep{Jaegle2021-fv}. 

Exceiver implicitly models gene expression as a discrete variable sampled from a count-based distribution. Global embeddings are used to represent gene identities, which are scaled by expression values upon model input. Exceiver also accounts for dropout, a sequencing bias characterized by experimental failure to capture the comprehensive set of genes expressed in an individual cell. This limitation implies that unobserved genes may be either truly unexpressed or merely unmeasured. Exceiver masks the attention computation at zero-valued expression positions to prevent learning based on this artifact. Upon decoding, Exceiver reuses unscaled gene embeddings in the output query matrix as a means to residually connect discrete gene identities within the learning process. Exceiver also provides an option to bias learning with an auxiliary classification task. An additional classification embedding token is concatenated to each input sequence and used to query the decoder. An auxiliary MLP classifies the output of the decoder from the classification token, and an additional cross-entropy loss is added to the mean squared error from the DNM task. Code is available at \url{https://github.com/keiserlab/exceiver}.

\subsection{Pretraining tasks}

Exceiver employs a self-supervised pretraining task as well as an optional auxiliary supervised classification task. The self-supervised task draws inspiration from masking tasks applied for NLP pretraining. However, there are several crucial differences between gene expression profiles and natural language. First, scRNA-seq features are a set rather than a sequence. Exceiver does not positionally encode gene embeddings, though features that describe spatial gene dependencies (such as the linear position of a gene or its relative position in the three-dimensional chromosome) may prove valuable in future work. Second, as previously referenced, expression features are counts of discrete features (genes), rather than identities of discrete features (words). As such, pretraining is a regression task rather than a classification task. Exceiver employs a new self-supervised task to account for the discrete distributions of features in the gene set.

The primary pretraining task we apply is discrete noise masking (DNM). First, we heuristically determine the number of genes to mask for each sample. We mask 15\% of the median number of recorded genes across the training dataset. We avoid masking a majority fraction of genes captured in poorly sampled cells by removing cells below a minimum number of recorded genes. In our experiments, we set this threshold at twice the number of masked genes. The training data therefore reflects a distribution of masked genes not exceeding 50\% of the observed genes in a given cell (Figure S2). A “mask” embedding replaces the randomly selected gene embeddings, and corresponding expression values are noised. Masked gene embeddings scaled by noised expression values pass through encoder and process modules. Finally, the embeddings of masked genes query the decoder. The resulting “contextual gene embeddings” pass through the final MLP, outputting expression predictions for masked genes (Figure 1a). 

Additionally, we propose an auxiliary classification task based on metadata labels. The Perceiver IO architecture is flexible enough to accommodate multimodal tasks. Exceiver implements an auxiliary classification task as described in the original paper. A classification token is initialized as a global embedding and passed through Exceiver with each gene set. This procedure allows for attention computation between all genes and the classification token. The original classification embedding queries the decoder along with masked genes, and the resulting “contextual classification embedding” is passed through an auxiliary MLP head, outputting a vector of class logits.

\subsection{Data and processing}

The Tabula Sapiens (TS) data was retrieved from figshare (\url{https://figshare.com/ndownloader/files/34702114}). Python packages anndata and scanpy were used for data storage and statistics calculations. Features were subsetted to all protein-coding genes as defined in the Cancer Cell Line Encyclopedia (CCLE) ($n=19067$). DecontX-corrected count matrices were used for training to account for ambient RNA. TS was shuffled and split into 70\% training and 30\% validation datasets. Counts were normalized to a maximum of $1e4$ per cell. Normalized counts were log-transformed and scikit-learn was used to apply a Z-score transformation to each dataset individually. Expression values were shifted to a mean-center of 1, implying the learned identity of global gene embeddings corresponds to average expression.

The Bi et al. data was retrieved from the Human Cell Atlas collection (\url{https://singlecell.broadinstitute.org/single_cell/study/SCP1288}). Raw count matrices were split into 70\% training and 30\% validation, and the same normalization and transformation procedure was applied. Missing genes were imputed with a value of zero.

The MIX-Seq data from McFarland et al. was retrieved from figshare (\url{https://figshare.com/s/139f64b495dea9d88c70}). Sanger GDSC2 AUC data was downloaded from the Dependency Map (\url{https://depmap.org/portal/download/custom/}). Drugs from 24-hr treatment experiments with sufficient matching GDSC2 experiments were selected for analysis. Post-perturbation transcriptome datasets were preprocessed individually in the manner as described previously with zero-value missing gene imputation. Cell line-drug transcriptome pairs were matched to dose-response curve AUC values of GDSC2 drug screening results.

\newpage
\section{Supplementary Figures}
\label{sup-figs}

\hfill
\hfill

\beginsupplement

\begin{figure}[!ht]
  \centering
  \includegraphics[width=0.7\textwidth]{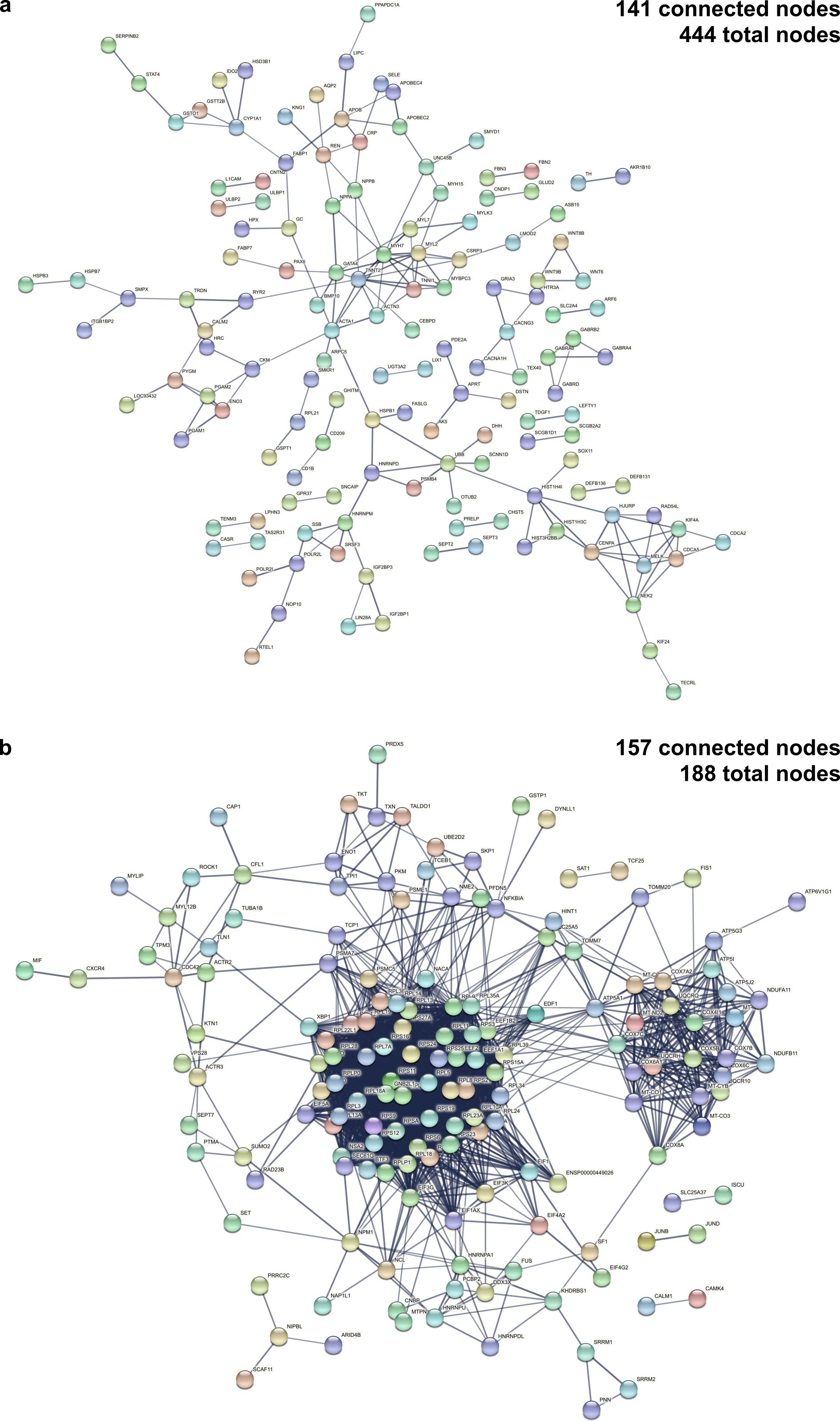}
  \caption{\textbf{Example clusters visualized using STRING.} Visualization of connected nodes from (a) cluster 17 and (b) cluster 45 from the STRING network of high confidence (interaction score > 0.7) interactions.}
  \label{sup-fig-1}
\end{figure}

\begin{figure}
  \centering
  \includegraphics[width=0.8\textwidth]{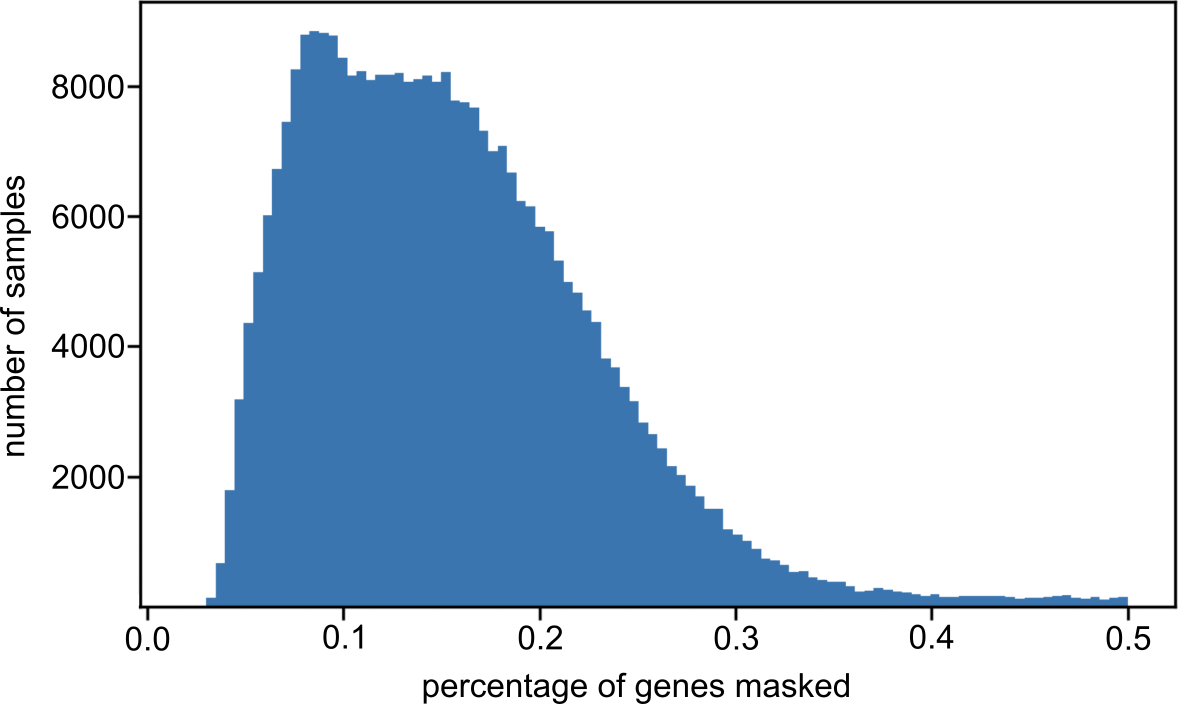}
  \caption{\textbf{Distribution of masked genes.} Histogram of the percentage of genes masked across the Tabula Sapiens training dataset. Feature sparsity results in a distribution of masked genes per sample.}
  \label{sup-fig-2}
  \vspace{400pt}
\end{figure}


\end{document}